\documentclass[reprint, amsmath, amssymb, aps, prb, longbibliography]{revtex4-2}
\usepackage{comment}
\usepackage{graphicx}
\usepackage{dcolumn}
\usepackage{bm}
\usepackage{ulem}
\usepackage{xcolor}

\begin{document}

\preprint{APS/123-QED}
\title{Gapfull and gapless $1$D Topological Superconductivity\\ in Spin-Orbit Coupled Bilayer Graphene}

\author{Daniel Skliannyi}\email{daniil.skliannyi@weizmann.ac.il}
\author{Yuval Oreg}
\author{Ady Stern}
\affiliation{Department of Condensed Matter Physics, Weizmann Institute of Science, Rehovot, Israel 76100}%

\date{\today}

\begin{abstract}

We propose a way to generate a one-dimensional topological superconductor from a monolayer of a transition metal dichalcogenide coupled to a Bernal-stacked bilayer of graphene under a displacement field. With proper gating, this structure may be tuned to form three parallel pads of superconductors creating two planar Josephson junctions in series, in which normal regions separate the superconductors. Two characteristics of the system which are essential for our discussion are spin orbit coupling induced by the transition metal dichalcogenides and the variation of the Fermi velocities along the Fermi surface. We demonstrate that these two characteristics lead to one-dimensional topological superconductivity occupying large parts in the parameter space defined by the two phase differences across the two junctions and the relative angle between the junctions and the lattice. An angle-shaped device in which this angle varies in space, combined with proper phase tuning, can lead to the formation of domain walls between topological and trivial phases, supporting a zero-energy Majorana mode, within the bulk of carefully designed devices. We derive the spectrum of the Andreev bound states and show that Ising spin-orbit coupling leaves the topological superconductor gapless, and the Rashba spin-orbit coupling opens a gap in its spectrum. Our analysis shows that the transition to a gapped topological state is a result of the band inversion of Andreev states.

\end{abstract}

\maketitle

\section{Introduction}
Topological superconductors have been a coveted goal in the last two decades, due to their unique physical properties and their potential applications for topological quantum information processing. 
Of particular interest have been one-dimensional (1D) topological superconductors, due to the localized Majorana zero modes that they carry at their ends. 
Several ways for engineering such superconductors were proposed, all combining superconductivity, spin-orbit coupling, and breaking of time-reversal symmetry. For most proposed ways superconductivity was induced by the proximity effect, spin-orbit coupling was intrinsic to the one-dimensional system, and time-reversal symmetry was broken by coupling a Zeeman field to the electron spin \cite{PhysRevLett.105.177002, PhysRevLett.105.077001, PhysRevB.85.020502,  PhysRevResearch.2.013377, PhysRevB.103.224505, PhysRevB.100.045301, doi:10.1126/science.aav3392, PhysRevB.84.085114, PhysRevB.84.014503,PhysRevB.89.115402, Aghaee2025, Zatelli2024, Flensberg2021}.   
\subsection{Double Josephson Junction}

In this work we consider a setup in which spin-orbit coupling is induced into the conducting system by the proximity of a transition metal dichalcogenide (TMD) layer to a Graphene bi-layer; superconductivity is intrinsic to the combined system  {(see Ref.~\cite{Zhang2023, zhang2024twistprogrammablesuperconductivityspinorbitcoupled})} or achieved by proximity  {(see Ref.~\cite{PhysRevB.100.035426})}; and time-reversal symmetry is broken by the application of phase differences across Josephson junctions, with or without the application of a Zeeman field. Furthermore, we introduce a new tuning knob, the relative angle $\beta$ between the Josephson junction and the underlying lattice, and show how it affects the formation of a topological superconductor.

\begin{figure}
\includegraphics[width=8.6cm]{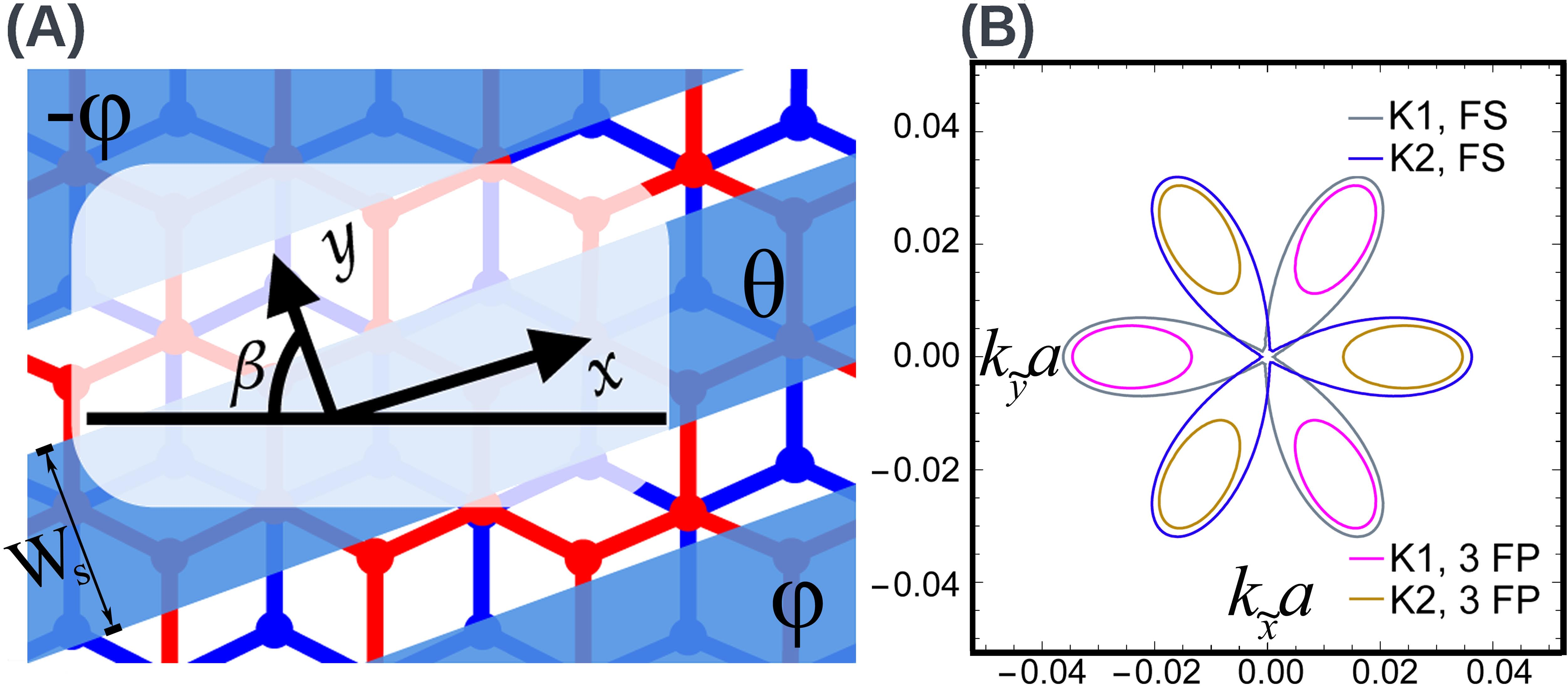}
\caption{
(A) A schematic of the system we consider, that consists of Bernal stacked bi-layer of graphene in proximity to TMD (which is not shown), on which patterned gates define superconducting (light blue) and normal (white) regions. The angle $\beta$ between the superconducting areas and the armchair orientation of the graphene bi-layer determines the velocity difference between the two Fermi branches. With an adjustment of the phase difference between the superconductors that form two Josephson junctions, we can bring the system into a topological state supporting Majorana zero modes.   
(B) Fermi surface of TMD/BBG heterostructure ($\beta=\pi/2$) in a perpendicular displacement field $D=1$ V$/$nm and $\mu=-12.653$ meV and $\mu=-12.2$ meV. For Fermi energy smaller than $-12.652$ meV we have three Fermi pockets (FP) per valley and for larger -- one Fermi surface (FS) per valley. The $k_{\tilde{x}}$ and $k_{\tilde{y}}$ are the coordinates in momentum space aligned with zig-zag and armchair directions, respectively. 
}\label{fig:warp}
\end{figure}

Our setup consists of a structure similar to Ref.\cite{PhysRevLett.131.146601}, comprising four layers: a bottom gate, a layer of transition metal dichalcogenide (TMD, e.g., WSe2), a Bernal stacking bilayer graphene (BBG), and a split top gate (See Fig.~\ref{fig:warp}(A)). In our case, the top gate is divided into three stripes. The angle between the stripes and the underlying Graphene lattice is $\beta$. 

Tuning the potentials of the bottom and top gates controls the density $n$ and the displacement field $D$ acting on the system. This allows one to adjust the system to achieve a superconducting state (S) or a normal state (N). The creation of two gaps in the top gate leads to the formation of an SNSNS structure that creates two Josephson junctions in series (we will later on name it $2$JJ), with the two external superconductors being semi-infinite and the middle superconductor having a width $W_{\text{s}}$. We note that the three stripes of the top gate may be replaced by superconducting pads, in which case the superconductivity is induced into the system by proximity rather than being intrinsic. 

Once the $2$JJ is generated, there are two additional important tuning parameters that control the system's properties - the two-phase differences between the three superconductors, represented by the phases $\theta$ and $\phi$, where $\theta$ is the phase of the order parameter on the central superconductor, and $\pm\phi$ are the phases of the external superconductors. 

 {\subsection{Discrete Vortex Condition}}

Earlier works, particularly Ref.\cite{PhysRevB.106.L241405, Lesser_2022, PhysRevB.103.L121116, doi:10.1073/pnas.2107377118}, found that when the electronic band structure leads to two branches with very different Fermi velocities $v_{{\rm f}s},v_{{\rm f}l}$  ($s,l$ stand for small and large) in the direction that crosses the junction, a 1D topological superconductor may be generated by tuning two phase differences across two Josephson junctions, with no need for a Zeeman magnetic field. Having a setup that is free of magnetic fields is desirable due to their adverse effects on superconductivity. A necessary condition for such a topological superconductor is that $\theta$ and $\phi$ form a discrete vortex (DV).  {Namely, the triangle formed by the three phases on the unit circle must enclose the centre of the circle, see Ref.~\cite{PhysRevB.90.155450}.}

\begin{figure}
\includegraphics[width=8.6cm]{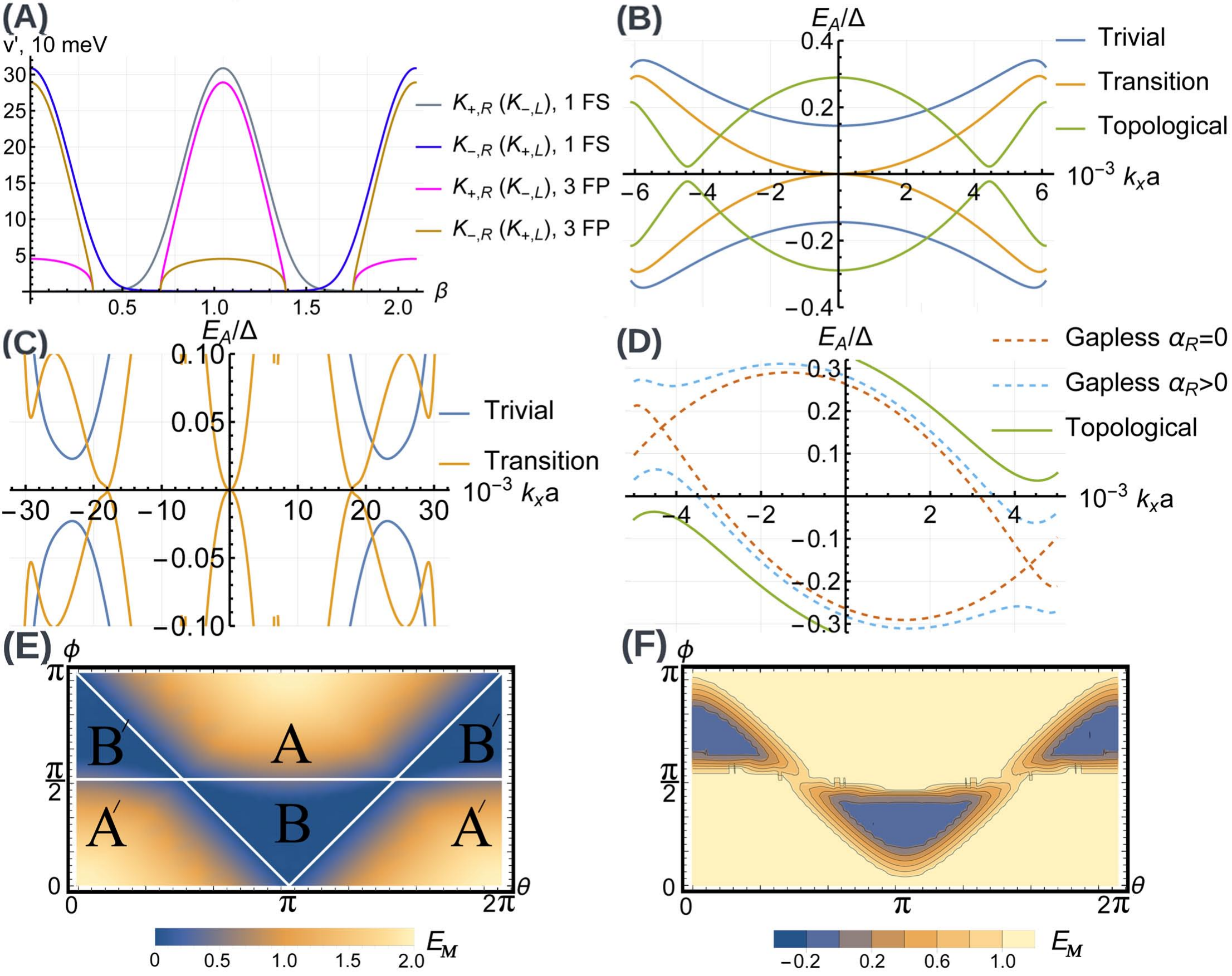}
\caption{(A) Fermi velocities (expressed in units of energy $v^{\prime}_{\rm f}=\hbar v_{\rm f} / a$, where $v_{\rm f}$ is Fermi velocity and $a=2.46 \text{ \r{A}}$. The energy scale  $300 \text{meV}$ corresponds to $\approx 10^5  \text{m} /\text{s}$) in different valleys ($K_{+}$ and $K_{-}$) as a function of angle $\beta$ $(k_{y}=0)$, for $\mu=-12.652$ meV and $\mu=-12.5$ meV. 
(B) and (C) The spectrum of $2$JJ for $\beta=0$ near the upper part (B) of the phase diagram and the lower part (C) for $\mu = -12.5$, $a=0.1$, $|d|=1$, ${\bf B}=0$ and $\Delta=0.17$. In panel (B), the conduction and the valence band of the lowest Andreev levels have opposite spins and the topological area is a consequence of band inversion of the Andreev spectrum. The value of $s_z$ of the negative energy band close to $E=0$ differs between neighboring topological triangles. Panel (C) shows that when Rashba coupling is added, left and right Fermi pockets can undergo a band inversion earlier than the middle Fermi pocket. (D) A deviation from high symmetry point $\beta=0$ to $\beta\neq0$, at which a threshold of Rashba coupling is required for the spectrum to be gapped. Panels (E) and (F) represent the minimal gap $E_{M}=Q\min_{k_{x},b}|E^{A}_{b}/\Delta|$ of Andreev state in $2$JJ over the phase space without (E) with (F) Rashba SO coupling, where $Q=-1$ in the topological area and $Q=1$ in trivial. White lines correspond to the maximal topological area according to the DV condition. $B$ and $B^{\prime}$ label topological gapless phases and $A$ and $A^{\prime}$ are gapped trivial phases related by spin flip.
}\label{fig:results}
\end{figure}

When the DV condition is fulfilled, the formation of a topological superconductor is easiest to understand in the limit where  $\xi_s \ll W_{\text{s}} \ll \xi_l$, where $W_{\text{s}}$ is the width of the middle superconductor and $\xi_s\propto v_{{\rm f}s}\text{ and }\xi_l\propto v_{{\rm f}l}$ are the coherence lengths for the two velocity branches, which are proportional to $v_{{\rm f}s},v_{{\rm f}l}$. In that limit, one branch, with coherence length $\xi_l$, is hardly affected by the middle superconductor and is insensitive to the phase of the middle superconductor, while the other, with coherence length $\xi_s$, experiences the system as two Josephson junctions in series and is, therefore, sensitive to both phase differences. This difference leads to single gap-closing lines in the $\phi,\theta$ parameter plane, which then leads to topological regions in that plane, see Fig.~\ref{fig:results}(F).

 {The area enclosed by these curves in the $\phi-\theta$ plane can be calculated as follows \cite{PhysRevB.106.L241405}:
\begin{equation}
\begin{split}
    S = 2\int_{\frac{\pi}{2}}^\pi d\phi \cos^{-1} &\left[ \tanh\left( \frac{W_{\text{s}}}{\xi_s} \right) \cos(\phi) \right]\\& - \cos^{-1} \left[ \tanh\left( \frac{W_{\text{s}}}{\xi_l} \right) \cos(\phi) \right] 
\end{split}
\end{equation}
Its maximum value is $ S_{\text{max}} = \frac{\pi^2}{4} $, which corresponds to $ \xi_s = 0 $ and $ \xi_l = \infty $.}

 {\subsection{Heterostructures Hamiltonian and Spin-Orbit coupling}}

 {Two key features of the BBG/TMD structure introduce unique characteristics into the system: its band structure and the predominance of Ising spin-orbit coupling. The low-energy band structure can be derived from the original Hamiltonian in the case $\beta = 0$:
\begin{widetext}
    \begin{equation}
H_{i}=\left(\begin{array}{cccc}
u/2 & v_{0}k_{\eta}^{\dagger} & -v_{4}k_{\eta}^{\dagger} & -v_{3}k_{\eta}\\
v_{0}k_{\eta} & f_{\text{BBG}}+u/2 & \gamma_{1} & -v_{4}k_{\eta}^{\dagger}\\
-v_{4}k_{\eta} & \gamma_{1} & f_{\text{BBG}}-u/2 & v_{0}k_{\eta}^{\dagger}\\
-v_{3}k_{\eta}^{\dagger} & -v_{4}k_{\eta} & v_{0}k & -u/2
\end{array}\right)\otimes s_{0}+\begin{pmatrix}\frac{\lambda_{I}}{2}\eta s_{z} + \frac{f_{\text{TMD}}}{2}s_{0} & \frac{\lambda_{R}}{2}\left(\eta s_{y}+is_{x}\right) & 0 & 0\\
\frac{\lambda_{R}}{2}\left(\eta s_{y}-is_{x}\right) & \frac{\lambda_{I}}{2}\eta s_{z} + \frac{f_{\text{TMD}}}{2}s_{0} & 0 & 0\\
0 & 0 & 0 & 0\\
0 & 0 & 0 & 0
\end{pmatrix},
\end{equation}
\end{widetext}
where the basis is defined as $\psi_{\eta}(k)=\left(\psi_{\eta,A_{1}}(k),\psi_{\eta,B_{1}}(k),\psi_{\eta,A_{2}}(k),\psi_{\eta,B_{2}}(k)\right)$, and $\eta=\pm1$ labels the valleys $\pm K=(\pm\frac{4\pi}{3a},0)$. The momentum dependence is given by $k_{\eta}=a(\eta k_{x}+ik_{y})$, where $a=0.24$ nm is the graphene lattice constant. The indices $A_{i}/B_{i}$ refer to sublattices, and $i=1,2$ denotes the two graphene layers. The matrices $s$ are Pauli matrices acting on the spin degrees of freedom. The parameter $f_{\text{TMD}}$ represents an internal displacement field arising from stacking, $\lambda_{I}$ is the Ising spin-orbit coupling constant, and $\lambda_{R}$ is the Rashba spin-orbit coupling constant. Both $\lambda_{I}$ and $\lambda_{R}$ are material-dependent and vary with the choice of TMD and its orientation relative to the graphene lattice (see Ref. \cite{PhysRevB.99.075438}).}

 {The parameters $v_{j} \equiv \frac{\sqrt{3}}{2} \gamma_{j}$, together with $\gamma_{1}$, $f_{\text{BBG}}$, and $u$, determine the band structure:
\begin{itemize}
\item $\gamma_{i}$ - various inter- and intra-layer hopping amplitudes;
\item $f_{\text{BBG}}$ - on-site potential difference due to stacking;
\item $u = -d_{\perp}D/\epsilon_{\mathrm{BBG}}$ - interlayer potential difference induced by an external displacement field, which opens a gap in the energy spectrum. Here, $d_{\perp}=0.33$ nm is the interlayer separation, and $\epsilon_{\mathrm{BBG}}\approx4.3$ is the relative permittivity of BBG.
\end{itemize}}

 {Achieving intrinsic conventional superconductivity in a TMD/BBG stack also requires the application of a displacement field (see Ref.~\cite{Zhang2023}). This requirement may originate from the dependence of spin-orbit coupling strength on the displacement field. As discussed in Ref.~\cite{Zhang2023, zhang2024twistprogrammablesuperconductivityspinorbitcoupled}, applying a perpendicular displacement field with the appropriate sign shifts the hole wave function closer to the TMD layer, thereby enhancing the spin–orbit coupling strength for the hole bands. Conversely, a displacement field of opposite sign moves the wave function away from the TMD, reducing the spin–orbit coupling for the given branch.}

 {If the chemical potential of the heterostructure does not lie within the range that supports intrinsic conventional superconductivity, one can instead rely on the superconducting proximity effect. Nevertheless, the displacement field still plays a significant role by influencing the spin–orbit coupling.}

 {To obtain the effective low-energy theory, we apply Brillouin-Wigner perturbation theory with respect to $\gamma_{1}$ \cite{10.1021/acs.nanolett.7b03604, PhysRevB.70.235111}, which simultaneously projects the full Hamiltonian onto the low-energy subspace, thereby reducing the Hilbert space dimension. The resulting effective Hamiltonian to leading and first order in $1/\gamma_{1}$ is given by:
\begin{equation}\label{Heff0}
    \mathcal{H}^{0}_{\text{eff}}=\left(
\begin{array}{cc}
 \frac{1}{2} (2w_{z}+ (f_{\text{TMD}}+u) s_{0}) & -k_{\eta} v_{3} s_{0}\\
 -k_{\eta}^{\dagger} v_{3} s_{0} & -\frac{1}{2} u s_{0} \\
\end{array}
\right),
\end{equation}
for zeroth order, and
\begin{widetext}
\begin{equation}
    \mathcal{H}^{1}_{\text{eff}}=\mathcal{H}^{0}_{\text{eff}}+\frac{1}{2\gamma_{1}}\left(
\begin{array}{cc}
 (v_{+}-v_{-}) ( |k_{\eta}|^{2} (v_{+}+v_{-})  s_{0}+ k_{\eta} a_{R,\eta}+k_{\eta}^{\dagger} a_{R,\eta}^{\dagger}) & -(v_{+}+v_{-}) \left( \frac{\left(v_{+}^2+v_{-}^2\right)}{\left(v_{+}+v_{-}\right)}(k_{\eta}^{\dagger})^{2}  s_{0}+k_{\eta}^{\dagger}a_{R,\eta} \right) \\
 -(v_{+}+v_{-}) \left( \frac{\left(v_{+}^2+v_{-}^2\right)}{\left(v_{+}+v_{-}\right)} (k_{\eta})^{2}  s_{0}+k_{\eta}a_{R,\eta}^{\dagger} \right) & (v_{+}-v_{-})|k_{\eta}|^{2}  \left(v_{+}+v_{-}\right)s_{0} \\
\end{array}
\right),
\end{equation}
\end{widetext}
for the first order, see Supplemental Material
  at [URL will be inserted by publisher] for the detailed derivation of the Hamiltonian. Here, we define $a_{R,\eta}=\frac{\lambda_{R}}{2}\left(\eta s_{y}+is_{x}\right)$, $v_{\pm}=v_{0}\pm v_{4}$, and $w_{z}=\frac{1}{2}\lambda_{I}s_{z}\eta$. The dominant role of the Ising spin-orbit coupling arises naturally from the perturbation expansion: $\lambda_{I}$ contributes already at zeroth order in $1/\gamma_{1}$, while Rashba terms appear only in higher orders.}

 {To account for a nonzero angle $\beta$, we perform a coordinate rotation:
\begin{equation}
    \begin{cases}
k_{\tilde{x}}\to\cos\left(\beta\right)k_{x}+\sin\left(\beta\right)(-i\partial_{y});\\
-i\partial_{\tilde{y}}\to-\sin\left(\beta\right)k_{x}+\cos\left(\beta\right)(-i\partial_{y}).
\end{cases}
\end{equation}
After the given rotation, one can find that the angle $\beta$ controls the degree of difference between the velocities of the two branches perpendicular to the junction.  Consequently, due to the highly anisotropic nature of the Fermi surface, $\beta$ can tune the system between topological and non-topological states.} The boundary between such regions contains a Majorana zero mode. Furthermore, we find that the effects of Ising and Rashba spin-orbit couplings on the phase diagram are different.
 {\subsection{Disorder and Topological Superconductivity}}
 {Disorder is an unavoidable and often significant factor in condensed matter systems, particularly in the study of topological phases. The effect of disorder on $ p$-wave superconductivity is a rich and extensively studied field. Early investigations focused on one-dimensional spinless $ p$-wave superconductors; see Refs.~\cite{PhysRevB.63.224204, PhysRevLett.107.196804, PhysRevLett.109.146403, PhysRevB.88.014206, PhysRevB.89.144506}. These studies show that the topological phase remains robust in the presence of weak disorder, but strong disorder ultimately drives a transition to a trivial phase. Interestingly, recent work~\cite{PhysRevB.100.035426} demonstrates that weak disorder at the interface between a nanowire and a proximitized superconductor can actually enhance the robustness of one-dimensional topological superconductivity by improving the superconducting proximity effect. Therefore, in engineered $ p$-wave superconductors, disorder can, in principle, enhance the stability of the topological phase, while still retaining its potentially detrimental effects.}

 {A similar dual behavior of disorder was observed in the case of topological superconductor–normal metal–superconductor Josephson junctions under an in-plane magnetic field (see Ref.~\cite{PhysRevLett.122.126801}). As shown in that work, in the presence of weak disorder, the region occupied by the topological phase in the parameter plane of superconducting phase difference and magnetic field is affected. In parallel, when the system is topological, weak disorder leads to a smaller localization length of the end Majorana modes.  As we discuss in subsection \ref{DS}, a similar picture holds also in our case. }

 {\section{Model Hamiltonian and gapless modes}}

The structure we consider,  BBG~\cite{McCann_2013} in a heterostructure with monolayers of TMD~\cite{Zhang2023, holleis2024nematicityorbitaldepairingsuperconducting}, carries two or four Fermi surfaces, one or two in each valley, which are $C_3$ symmetric, but are anisotropic due to trigonal warping, see Fig.~\ref{fig:warp}(B). It is 
extensively studied both theoretically \cite{McCann_2013, 10.1021/acs.nanolett.7b03604, PhysRevLett.131.146601} and experimentally\cite{Zhang2023, holleis2024nematicityorbitaldepairingsuperconducting}. 

Generally, the Hamiltonian of the $2$JJ system that we consider is described in terms of valleys $\eta_z=\pm 1$, spins $s_z=\pm 1$ and particle-hole $\tau_z=\pm 1$. We use a convention in which the electron spin direction is $\tau_zs_z$. In this convention, $s$-wave pairing is $s$-independent. The Hamiltonian then is, 
\begin{equation} \label{hamiltonian1}
\begin{split}
    H = \Big[\epsilon(\eta_z {\bf k}) - \mu & + \alpha_I \eta_z s_z  
    + \alpha_R {\bf k} \times {\bf s} \Big] \tau_z  \\
    &+ {\bf B}\cdot{\bf s} + \Delta_x(y) \tau_x + \Delta_y(y) \tau_y,
\end{split}
\end{equation}
where $\epsilon(k_x,k_y)$ is the band energy, $\alpha_I$ is the Ising spin-orbit coupling constant, $\alpha_R$ is the Rashba spin-orbit constant, ${\bf B}=(B_{x},B_{y},0)^T$ is an in-plane magnetic field, $\bf k$ is measured from the center of the valley, and $\Delta_{x,y}(y)$ is superconducting order parameter that defines the $2$JJ. Notice that since the $x,y$ directions are determined by $\beta$, the dependence of $\epsilon$ on $\bf k$ is rotated when  $\beta$ is varied, Fig.~\ref{fig:warp}.

Assuming the junction is translation invariant in the $x$-direction, $k_x$ is a good quantum number and can be considered as a parameter, we can state that the Hamiltonian (\ref{hamiltonian1}) is of class D, for which states may be classified as either trivial or topological. A transition between trivial and topological states occurs when there is a single gap closing at $k_x=0$ \cite{PhysRevX.7.021032, PhysRevLett.118.107701}. 

We analyze the Hamiltonian (\ref{hamiltonian1}) in two complementary ways. In the first part, we employ symmetry arguments and approximate solutions to examine the properties of the spectrum of the two Andreev modes that are closest to zero energy. In the second part, we evaluate eigenvalues and study them for the full Hamiltonian analytically, using a symmetry argument to select appropriate couplings. In both parts, we separate between the cases of $\beta=0$ and $\beta\ne 0$. For each of these two cases we study the effects of Ising coupling, magnetic field, and Rashba coupling on the junction band structure and topology.

As we now show, when the spin-orbit coupling is entirely of the Ising type, the topological region is always gapless due to the gap being closed at $k_x\ne 0$, and becomes gapped only when Rashba spin-orbit coupling is introduced. Furthermore, we find that due to the structure of the Fermi surface, the phase diagram in the $\phi,\theta$ plane strongly depends on $\beta$, allowing for Majorana modes to be placed in regions where $\beta$ is changed.

{\it We first focus on the case $\beta=0$ and search for solution of linearized BdG equation.} 
The low-energy Andreev mode $E_b(k_x)$ (with $b$ denoting the velocity branch) may be analytically derived in the absence of Rashba coupling and a magnetic field, by linearizing the kinetic term in the BdG equations. In the limit in which the normal regions are wide, such that there are many sub-gap Andreev states for each $k_x$, the linearized BdG equation yields
\begin{equation}
   \cos\left(\frac{E_{p,b}(k_{x})}{E^{T}_{n,b}}-\phi\right) = -\cos (\theta) R_{b},
   \label{dispersionlong}
\end{equation}
where $p$ is an integer that enumerates the mode, 
\begin{equation*}\label{AR}
R_{b} =\max\left(\tanh \left(\frac{\Delta}{E^{T}_{sc,b}}\right) - \Gamma,0\right),
\end{equation*}
 $b=s,l$ label the velocity branches and $\Gamma$ is a phenomenological suppression factor of the Andreev reflection amplitude $R_b\ge 0$.  The Thouless energy for the superconducting region is:
\begin{equation*}
E^{T}_{sc,b}(k_{x})=\frac{v_{b{\rm f}}(k_{x})}{ W_{\text{s}}},
\end{equation*}
where $W_{\text{s}}$ is the width of the superconducting region. { The energy $E^{T}_{n,b}(k_{x})=\frac{v_{b{\rm f}}(k_{x})}{ W_{n,\text{l}}+W_{n,\text{r}}}$ is defined for the normal region in a similar way, where $W_{n,\text{l/r}}$ are width of left and right normal metal respectively}, see Supplemental Material
  at [URL will be inserted by publisher] for the detailed derivation of the Andreev energies.

 {Since mirror symmetry with respect to the $x$-direction is already broken by the phase difference and the widths of the normal regions on either side contribute additively, one can also observe that precise fine-tuning of these widths is not essential for achieving a topological state. Furthermore, in contrast to Ref. \cite{PhysRevX.7.021032}, the nontrivial phase of the central superconductor prevents the system from re-entering a time-reversal symmetric phase. As a result, the system remains in symmetry class D without requiring distinct superconducting gaps in the left and right leads.}

In the limit where the normal regions are narrow ($E_T^n\gg \Delta$) we find a single set of in-gap modes:
\begin{equation}
    \frac{E_{b}}{\Delta}= \pm\cos\frac{\phi\pm \arccos\left(\frac{2 \cos\left(\theta\right) R_{b}+ \cos\left(\phi\right)\left(1-R_{b}\right)}{\left(1+R_{b}\right)}\right)}{2}.
    \label{dispersionshort}
\end{equation}

Gapless states correspond to $E_{b}=0$. As can be seen in the expressions for $E_{b}$, when a gapless solution exists for some particular values of $\theta,\phi,k_x$ it may still exist around these values because the changes in $\theta,\phi$ may be compensated by a change in $R_b$, which is determined by $k_x$. Gap closings may then disappear or be created only in pairs, either at $k_x=0$, in which case they correspond to a topological transition, or at $k_x\ne 0$. This stability of the zero energy states may be understood from symmetry considerations, as we now discuss. 

\section{Topological (Nodal) Superconductivity}

For $\beta=0$ the Hamiltonian (\ref{hamiltonian1}) anti-commutes with the Particle-hole anti-unitary operator ${\hat P}=K\eta_x s_y\tau_y$, with $K$ being complex conjugation that flips the sign of the momentum. In the absence of Rashba coupling,  the Hamiltonian anti-commutes also with a particle-hole operator that is local in $k_x$, given by ${\hat P}_{k_x}={\tilde K}\eta_x s_y\tau_y$, where $\tilde K$ is a complex conjugation that leaves $k_x$  intact.  As we now show, when either $\{{\hat P}_{k_x}, H\}=0$ or $[H,s_z]=0$ the topological region is gapless, and hence Rashba spin-orbit coupling is essential for a gap to open in that region. 

When $\{H,{\hat P}_{k_x}\}=0$, the spectrum of $H(k_x)$ is symmetric around zero energy. When there are two zero energy states at $k_x=k_0$, close to $k_0$ the Hamiltonian may be written in the two-dimensional Hilbert space spanned by these two states. All Hermitian operators may then be described in terms of a set of Pauli matrices $\rho_i$ with $i=0, \dots,3$ acting within this subspace, and the operator ${\hat P}_{k_x}$ may be represented in that subspace as the product of complex conjugation and a Pauli matrix $\rho_x$. The only form of the Hamiltonian allowed by the condition $\{H, \tilde{K}\rho_x\}=0$ is then $h\rho_z$, with $h$ being a real function of $\delta k_x =k_x-k_0$ and the other system parameters, in particular $\phi,\theta$ and the magnetic field $\bf B$. The two states are degenerate when $h=0$. Thus, a small change in $\phi,\theta, {\bf B}$ may be compensated by a small change in $\delta k$ to keep $h=0$, rendering the junction's spectrum gapless. This is indeed reflected in the energies from Eq.~(\ref{dispersionlong}) and from Eq.~(\ref{dispersionshort}), in which a small change in $\theta,\phi$ may be compensated by a small change in $R^{s/l}$, which depends on $k_x$. Moreover, this gapless area is the area at which the phases form a discrete vortex. The topological nature of this vortex allows us to define this gapless state as a topological nodal superconductor.

A similar state of matter can also arise in a $2$D Ising superconductor when an in-plane magnetic field exceeds the superconducting gap, see Ref.~\cite{10.21468/SciPostPhys.12.6.197, He2018, PhysRevB.109.165427}. In our case the nodal superconductivity arises without application of in-plane magnetic field. The application of an in-plane magnetic field, together with potential barriers or disorder, opens a gap in the spectrum, giving rise to a gapped topological phase. Notably, the underlying physics in both situations has a similar topological origin: nodal points emerge as topological phase transitions between trivial and topological regions in momentum space.


 \subsection{Topological Periodic Table, Principle I}
 
The gapless phase we study here possesses D-class symmetries in $d=1$. It is a 1D analog to the Weyl semimetal (WS) phase that occurs in 3D. Similar to our discussion, the existence of the WS phase ~\cite{RevModPhys.90.015001} is argued by fixing one momentum component, say $k_z$, in the 3D Brillouin zone and evaluating the topological invariant of the resulting 2D system as a function of $k_z$. A change of that invariant requires a gap closure, which forms a Weyl cone. Our case for $\beta=0$ is similar to that of a WS phase in which the system possesses a crystalline symmetry that makes the $d=2$ Hamiltonian invariant to  $k_z\rightarrow -k_z$, such that the $d=2$ system of fixed $k_z$ has the same symmetry class of its $d=3$ parent phase. Motivated by this observation, we propose the following general principle regarding the periodic table of topological insulators and superconductors: If a topological symmetry class contains a sequence of dimensions $d_{1}=d_{2}-1$ characterized both by $\mathbb{Z}_{2}$ invariants, it becomes possible, in the presence of an additional non-local symmetry, to realize a topologically gapless phase in dimension $d_{2}$. 

\begin{figure}
\includegraphics[width=8.6cm]{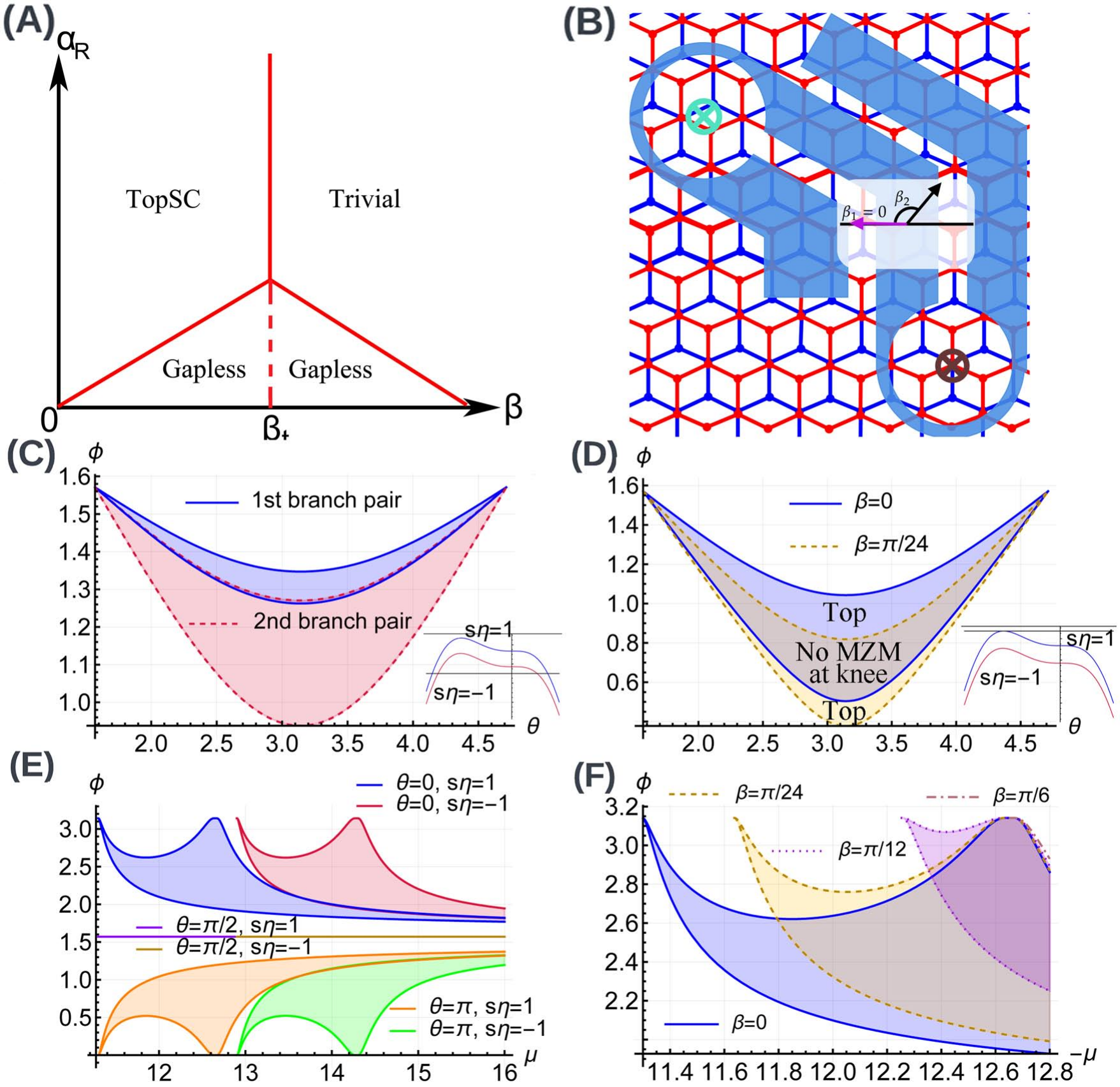}
\caption{(A) $\beta - \alpha_R$ qualitative phase diagram of a $2$JJ. The phases of the superconductors are fixed in the middle of a topological region. The diagram shows two gapped regions (Trivial and Topological) and a gapless region. The gapless region connects between gapped trivial and topological superconductors. The angle $\beta_t$ represents the alignment angle of the device with respect to the tangential direction of the Fermi pockets that separates the trivial and the topological states. (B) A possible configuration for a $2$JJ device oriented on the lattice of BBG in such a way that, for some value of $\theta$ and $\phi$, it acquires a trivial region in one half of the device and topological region at another, with Majorana zero modes between them. The $\beta_{1}=0$ and $\beta_{2}$ are corresponding angles, and color crosses are phase-controlling fluxes. (C) Phase diagram for the case of four velocity branches,  for $\mu=-15$ meV and $\beta=0$. The colored regions represent gapped topological phases. The diagram shows that the topological region can expand when the system has two bands per valley instead of one. (D) Phase diagram for $\beta_{1}=0$ (the shaded blue region is topological) and $\beta_{2}=\pi/24$ (the shaded gold region is topological). The chemical potential is $\mu=-12$ meV. In a knee geometry with angle $\pi/24$, when both parts are topological (the blue and gold regions overlap), there is no MZM at the knee. When only one of them is topological, the knee hosts an MZM. { (E) $\mu$–$\phi$ phase diagram for different values of $\theta$. One can clearly observe the strong dependence of the phase diagram on the chemical potential, as well as the reentrance of the topological phase as $\mu$ is lowered. (F) $\mu$–$\phi$ phase diagram for fixed $\theta=0$ and different values of $\beta$. } }\label{fig:domainwall}
\end{figure}

 {\subsection{ Andreev Band Inversion}}
In the absence of ${\hat P}_{k_x}$ breaking terms, the spectrum may be gapped only when the Hilbert space of nearly zero-energy states is larger than two-dimensional, e.g., four-dimensional. This happens in general when zero-energy states at two different values of $k_x$ move toward one another. In particular, it happens when $k_0$ approaches zero, since for $\beta=0$ the spectrum is even with respect to $k_x$. The spectrum may be gapped also by having a term that breaks the relation $\{H,{\hat P}_{k_x}\}=0$. Examples to this are Rashba spin-orbit coupling and a space-dependent magnetic field with a wave vector that couples the two valleys and leads to a term $\textbf{B}  \cdot \vec s  \eta_x$. When the gapping happens with two gapless points that merge at $k_0=0$, the transition between trivial and topological superconductor is a band inversion of two Andreev bands, see Fig.~\ref{fig:results}(B).

This general structure may be understood in another way. When the normal regions between the superconductors are small, there are eight sub-gap energy modes. We will focus on the four lowest among them, which are described by the energies in Eq.~(\ref{dispersionshort}) with a plus sign before the $\arccos$ function. The subgap spectrum may then be described in terms of a $4\times 4$ Hamiltonian, which we span by two sets of Pauli matrices $\lambda\text{ and }s$, where $s$ is the same as that appearing in Eq.~(\ref{hamiltonian1}) while $\lambda$ acts on the particle-hole space and is related to $\tau$ by a unitary transformation that is spin-independent. We write the diagonalized $4\times 4$ Hamiltonian as $H=\frac{E_A^{l}+E_A^s}{2}\lambda_{z}+\frac{E_A^{l}-E_A^s}{2}s_{z}$. This Hamiltonian anti-commutes with the particle-hole operator, which may be chosen here to be $K\lambda_y s_y$. At the transition, then, the two states of zero energy differ both by their $\lambda_z$ and $s_z$ values, i.e., they have the same eigenvalue of $\lambda_z s_z$. For an operator to couple these two states and open a gap, it must therefore flip both $\lambda_z$ and $s_z$. However, for such an operator to flip both of these numbers and still anti-commute with the particle-hole operator, it must be odd under $k_x\rightarrow -k_x$.

 {\subsection{Various Perturbations and Disorder}}\label{DS}

These observations allow us to analyze the effect of various perturbations on the low-energy Andreev modes. The energy modes in Eq.~(\ref{dispersionshort}) are calculated by linearizing the BdG equations, thus setting normal reflections to zero, and by setting the magnetic field and the Rashba spin-orbit coupling to zero as well. Normal reflection, which can arise at the junction border, does not flip the spin, and will therefore introduce terms of the form $V_x\lambda_x+V_y\lambda_y$. These terms would not couple the two states close to zero energy, and would therefore not open a gap. An in-plane Zeeman field will introduce terms of the form $b_x s_x+b_y s_y$. Again, these terms will not open a gap, for the same reason. Finally, Rashba spin-orbit coupling, being odd with respect to $k_x\rightarrow -k_x$, will be able to couple two states of the same $\lambda_z s_z$, and will therefore open a gap that will have band inversion. Moreover, the interplay between potential-induced scattering and either weak Rashba spin-orbit coupling or a weak magnetic field can, perhaps surprisingly, drive the system into a topological gapped phase.

 {The effects of weak disorder, prior to disorder averaging, can be incorporated by replacing the couplings discussed above with random variables. In particular, once projected to the four sub-bands space we consider, potential disorder gives rise to nontrivially correlated couplings of the form  
\begin{equation}
V^{D}_x(x)\lambda_x + V^{D}_y(x)\lambda_y .
\end{equation}
This noise can, for example, be modeled as delta-correlated with the structure  
\begin{equation}
\langle V^{D}(x) \left(V^{D}(x')\right)^{\dagger} \rangle = V_0 \, \delta(x - x'),
\end{equation}
where $ V_0 $ is a constant and $ V^{D}(x) = V^{D}_x(x) + i V^{D}_y(x) $.  
The resulting energy spectrum and density of states are then obtained from a BdG-like Hamiltonian, with $ k_{x} $ replaced by $ -i\partial_{x} $.}

 {The effect of disorder can already be understood at the level of the unaveraged Hamiltonian. These random coupling terms induce level repulsion between same-spin particle-hole--related states, thereby suppressing the topological phase. As a result, the topological region narrows with increasing disorder strength, although it persists up to a critical value. It can be understood from Fig.~\ref{fig:results}(D, $\alpha_{R}=0$). Potential disorder, in case of $\alpha_{R}=0$, will couple one of the low energy levels with the high energy level with opposite particle-hole Pauli matrix value $\lambda_{z}\to-\lambda_{z}$ and same spin. Such coupling will suppress the band-inverted part of the spectrum and return the system to a trivial, straight-gap state.}

 {More intriguing phenomena can arise from the interplay between weak disorder and Rashba spin-orbit coupling. This combination can, in principle, increase the localization length of Majorana zero modes \cite{PhysRevLett.122.126801}, which corresponds to an enhancement of the minimal gap in the gapped topological phase for translationally invariant barriers. Similarly, for $\beta\neq0$, it can drive a transition from a gapless to a gapped topological superconducting state, which corresponds to an effective enhancement of the $\alpha_{R}$ coupling strength, see in Fig.~\ref{fig:results}(D).}

 {\section{$y$-direction Mirror symmetry breaking}}

In the case of $\beta\neq0$, the Hamiltonian does not anti-commute with ${\hat P}_{k_x}$ and becomes asymmetric with $k_{x}\to-k_{x}$ transformation, leading to breaking of mirror symmetry concerning the $y$ axis. However, in the absence of Rashba coupling and Zeeman field, it commutes with $s_z$. For every $\beta$ and for any $k_x$ Eq.~(\ref{dispersionshort}) still holds, but the dispersion is not symmetric with respect to $k_x\rightarrow -k_x$. Thus, crossing points of the modes with two opposite values of $ s_z$ do not necessarily occur at zero energy. Close to these points the effective Hamiltonian may be written as $E_0+\alpha \delta k_{x} s_z$, where $\delta k_{x}=k_{x}\mp k_0$, and there are then four crossing points of the zero energy line, again making the topological region gapless. Furthermore, in this case, infinitesimal Rashba coupling removes the degeneracy at $E_0$, but does not gap the spectrum at zero energy, necessitating a threshold value of the Rashba coupling to introduce a gapped topological region.  A schematic plot of the resulting phase diagram in the $\beta$-$\alpha_R$  plane is given in Fig.~\ref{fig:domainwall}(A).

 {\subsection{Topological Periodic Table, Principle II}}
When $\beta\ne 0$ the zero dimensional system defined when $k_x$ is fixed has no symmetry, and thus belongs to class A, which follows a topological classification of $\mathbb{Z}$. Again, transitions between different values of the topological invariant as a function of $k_x$ result in gap closures and gapless 1D superconductivity. Since for class A the spectrum is not symmetric around zero, single gap crossings are allowed. More generally, then, we see that if in the sequence of dimension $d_{1}=d_2-1$ the $d_{1}$ dimension corresponds to a symmetry class with $\mathbb{Z}$ invariant and $d_{2}$ corresponds to a symmetry class with $\mathbb{Z}_{2}$ invariant it becomes possible to realize a topological gapless phase in $d_{2}$ dimension.

 {\subsection{Domain Walls}}
The $\beta$-dependence of the phase diagram may be used in geometries in which $\beta$ varies along the junction. For example, Fig.~\ref{fig:domainwall}(B) shows a knee-like 2JJ, where $\beta$ varies on the two sides of the knee by an obtuse angle, and where each end of the junction has a phase controlling loop, to control $\theta$ and $\phi$. The variation of $\beta$ gives rise to different phase diagrams for each orientation, see Fig~\ref{fig:domainwall}(D). This geometry creates a spatial mismatch of $2$JJ topological properties and, for some particular $\theta$ and $\phi$, gives rise to a topological domain wall that supports a Majorana zero mode at the knee itself.

 {\subsection{Band Filling and Reentrance of Topological Superconductivity}}

One can employ the Hamiltonian from Eq.~(\ref{Heff0}) to compute the phase diagram as a function of $\phi$, $\theta$, $\beta$ and the chemical potential $\mu$. The resulting phase diagrams are shown in Fig.~\ref{fig:domainwall}(C-E), which describe how the gapped topological region behaves with changes in the chemical potential. When the chemical potential is such that two bands are being filled in each valley (see the inset to Fig.~\ref{fig:domainwall}(C) for the band structure in one valley), the topological region expands. In the absence of normal reflection, each of the bands forms a phase diagram of its own as a function of $\theta,\phi$. In regions where both bands form topological superconductors, normal reflection may turn both to be trivial, in view of the $Z_2$ nature of the classification of Class D.

 {The full dependence of the phase diagram on $\mu$ for fixed $\theta$ is shown in Fig.~\ref{fig:domainwall}(E) and Fig.~\ref{fig:domainwall}(F) (Notice that scales of the x-axis are different between the panels). As seen from these diagrams, the topological region initially expands as the chemical potential decreases, reaching a maximum around $\mu = -12.63$ meV. Beyond this point, the topological area begins to shrink, with the shrinking continuing until the onset of filling of the next band. Interestingly, the filling of the second band leads to a reemergence of a large topological region, as the behaviour observed with the first band. This can be understood by noting that the Ising spin–orbit coupling merely shifts the spin-polarised bands in one valley without altering their dispersion.}

 {\section{Summary}}

{To summarize}, we studied a 2JJ geometry made of a two-dimensional structure of Bernal bi-layer of graphene tunnel coupled to a monolayer of a TMD material. We found that when the spin-orbit coupling induced by the TMD in the graphene bilayer is purely of the Ising type, a region of topological superconductivity exists in the phase diagram, but the topological region is gapless.
Spin-rotating couplings, such as Rashba coupling, is crucial for the existence of a nonzero gap in the topological region. We also showed that the relative direction of the junction with respect to the lattice determines the fraction of the phase diagram that is occupied by a topological state.

\section{Acknowledgment}

We thank Aleksandr Osin, Vishal Bhardwaj, Omri Lesser, Yuli Nazarov, Mikhail Feigel'man, Boris Altshuler and Alexei Kitaev for their helpful discussions. D.S. acknowledges the Ljubljana Summer School on Quantum Physics, which was held at the University of Ljubljana, the Jozef Stefan Institute, and the CENN Nanocenter. This work was supported by the Israeli Science Foundation, the Israeli Ministry of Science Technology and Space, the Minerva Stiftung, the European Union’s Horizon 2020 research and innovation programme (Grant Agreement LEGOTOP No. 788715), the DFG (CRC/Transregio 183, EI 519/7-1), the Israel Science Foundation ISF (Grant No 1914/24) and ISF Quantum Science and Technology (2074/19).
\nocite{*}

\bibliography{apssamp}

\appendix
\begin{widetext}
\section*{Supplemental Material}
\subsection*{S.1 Explicit derivation of perturbative effective Hamiltonian.}
\subsubsection*{BBG tights binding model}
\label{clear}
We begin with the tight-binding Hamiltonian for bilayer graphene (BBG) under a perpendicular displacement field $D$:  
\begin{equation}
H_{\mathrm{BBG}}=\left(\begin{array}{cccc}
u/2 & v_{0}k_{\eta}^{\dagger} & -v_{4}k_{\eta}^{\dagger} & -v_{3}k_{\eta}\\
v_{0}k_{\eta} & f_{\text{BBG}}+u/2 & \gamma_{1} & -v_{4}k_{\eta}^{\dagger}\\
-v_{4}k_{\eta} & \gamma_{1} & f_{\text{BBG}}-u/2 & v_{0}k_{\eta}^{\dagger}\\
-v_{3}k_{\eta}^{\dagger} & -v_{4}k_{\eta} & v_{0}k_{\eta} & -u/2
\end{array}\right).
\end{equation}

Our basis is defined as $\psi_{\eta}(k)=\left(\psi_{\eta,A_{1}}(k),\psi_{\eta,B_{1}}(k),\psi_{\eta,A_{2}}(k),\psi_{\eta,B_{2}}(k)\right)$,  
where $\eta=\pm1$ corresponds to the valleys $\pm K=(\pm\frac{4\pi}{3a},0)$. The term $k_{\eta}=a(\eta k_{x}+ik_{y})$ represents the momentum dependence of the BBG Hamiltonian, with $a=0.24$ nm being the graphene lattice constant. The indices $A_{i}/B_{i}$ denote sublattices, while $i=1,2$ labels the two layers.  

The parameters $v_{j}\equiv\frac{\sqrt{3}}{2}\gamma_{j}$, along with $\gamma_{1}$, $f_{\text{BBG}}$, and $u$, define the band structure:  
\begin{itemize}
    \item $\gamma_{i}$ – various hopping amplitudes;  
    \item $f_{\text{BBG}}$ – on-site potential difference arising from stacking;  
    \item $u=-d_{\perp}D/\epsilon_{\mathrm{BBG}}$ – interlayer potential difference induced by the displacement field, where $d_{\perp}=0.33$ nm is the interlayer separation and $\epsilon_{\mathrm{BBG}}\approx4.3$ is the relative permittivity of BBG.  
\end{itemize}  

From \cite{PhysRevB.89.035405}, we adopt the \textit{ab initio} calculated parameters: the intralayer nearest-neighbor hopping is $\gamma_{0}=2.61$ eV, while the interlayer hopping amplitudes are $\gamma_{1}=361$ meV, $\gamma_{3}=283$ meV, and $\gamma_{4}=138$ meV. The on-site potential difference is $f_{\text{BBG}}=15$ meV.  

This Hamiltonian differs from that used in \cite{10.1021/acs.nanolett.7b03604}, where a simplified version was considered, specifically setting $v_{3}=v_{4}=0$.

\subsubsection*{Adding of transition metal dichalcogenides top layer}\label{Heff}
We can treat the addition of the TMD top layer like a perturbation that affects the top layer of BBG; namely, we take into account only virtual transitions between the top monolayer graphene (MLG) and TMD material. Notice that the chemical potential is always in the gap of the TMD. In this case, we can project the Hamiltonian on the BBG bands. Transitions take place only from the BBG top layer to the TMD and lead to the following spin-dependent Hamiltonian:
\begin{equation}
H_{\mathrm{SOC}}(k)=\mathcal{P}_{1}\delta H_{\mathrm{MLG}}\mathcal{P}_{1},
\end{equation}
where $\mathcal{P}_{1}=\left(\begin{array}{cccc}
1 & 0 & 0 & 0\\
0 & 1 & 0 & 0\\
0 & 0 & 0 & 0\\
0 & 0 & 0 & 0
\end{array}\right)$ -- projection operator in basis mentioned above and 
\begin{equation}
\delta H_{\mathrm{MLG}}= \frac{f_{\text{TMD}}}{2}\sigma_{z} s_{0} + \frac{\lambda_{I}}{2}\eta \sigma_{0}s_{z}+\frac{\lambda_{R}}{2}\left(\eta\sigma_{x}s_{y}-\sigma_{y}s_{x}\right)
\end{equation}
-- proximity contribution of $\textrm{WSe}_{2}$/MLG Hamiltonian. Where in MLG Hamiltonian $\sigma$ and $s$ is Pauli matrix that act on
sublattice $(A, B)$ and spin degrees of freedom respectively, $f_{\text{TMD}}$ -- internal displacement field that arise due to stacking, $\lambda_{I}$ is rising spin-orbit coupling constant and $\lambda_{R}$ -- Rashba coupling constant. It was described in work \cite{PhysRevB.92.155403} and added spin-orbit coupling potential that arises from the coupling between BBG and TMD material after stacking.

Full effective Hamiltonian, in this case, will be:

\begin{equation}
H_{\text{eff }}=H_{\mathrm{BBG}}\otimes s_{0}+\mathcal{P}_{1}\delta H_{\mathrm{MLG}}\mathcal{P}_{1},
\end{equation}

In addition to all of the above, one can note that the spin-orbit coupling constants in the Hamiltonian depend on the rotation angle of the TMD sheet relative to the BBG. The values of these parameters are obtained in paper \cite{PhysRevB.99.075438} in the case of MLG and can be extrapolated to the case considered in this article.

\subsubsection*{Brillouin-Wigner perturbation theory}
It is well-known, that for generic momentum near K points the strongest hopping in BBG occurs between the B1 and A2 sites -- $\gamma_{1}$, hence we will treat:
\begin{equation}
H_{\gamma_{1}}=\left(\begin{array}{cccc}
0 & 0 & 0 & 0\\
0 & 0 & \gamma_{1} & 0\\
0 & \gamma_{1} & 0 & 0\\
0 & 0 & 0 & 0
\end{array}\right)\otimes s_{0},
\end{equation}
like initial Hamiltonian and all other terms: 

\begin{equation}
H^{'}=\left(\begin{array}{cccc}
u/2 & v_{0}k_{\eta}^{\dagger} & -v_{4}k_{\eta}^{\dagger} & -v_{3}k_{\eta}\\
v_{0}k_{\eta} & f_{\text{BBG}}+u/2 & 0 & -v_{4}k_{\eta}^{\dagger}\\
-v_{4}k_{\eta} & 0 & f_{\text{BBG}}-u/2 & v_{0}k_{\eta}^{\dagger}\\
-v_{3}k_{\eta}^{\dagger} & -v_{4}k_{\eta} & v_{0}k & -u/2
\end{array}\right)\otimes s_{0}+\begin{pmatrix}\frac{\lambda_{I}}{2}\eta s_{z} + \frac{f_{\text{TMD}}}{2}s_{0} & \frac{\lambda_{R}}{2}\left(\eta s_{y}+is_{x}\right) & 0 & 0\\
\frac{\lambda_{R}}{2}\left(\eta s_{y}-is_{x}\right) & \frac{\lambda_{I}}{2}\eta s_{z} + \frac{f_{\text{TMD}}}{2}s_{0} & 0 & 0\\
0 & 0 & 0 & 0\\
0 & 0 & 0 & 0
\end{pmatrix},
\end{equation}
as a perturbation.

Initial Hamiltonian and perturbation in new basis, where $H_{\gamma_{1}}$ is diagonal, have the next form:

\begin{equation}
H_{\gamma_{1}}=\left(\begin{array}{cccc}
0 & 0 & 0 & 0\\
0 & 0 & 0 & 0\\
0 & 0 & -\gamma_{1} & 0\\
0 & 0 & 0 & \gamma_{1}
\end{array}\right)=\gamma_{1}\left(\begin{array}{cccc}
0 & 0 & 0 & 0\\
0 & 0 & 0 & 0\\
0 & 0 & -1 & 0\\
0 & 0 & 0 & 1
\end{array}\right),
\end{equation}
\begin{equation}
H^{'}=\left(\begin{array}{cccc}
\frac{1}{2}(2w_{z}+(u + f_{\text{TMD}})s_{0}) & -k_{\eta}v_{3}s_{0} & \frac{k_{\eta}^{\dagger}v_{+}s_{0}+a_{R,\eta}}{\sqrt{2}} & \frac{k_{\eta}^{\dagger}v_{-}s_{0}+a_{R,\eta}}{\sqrt{2}}\\
-k_{\eta}^{\dagger}v_{3}s_{0} & -\frac{u}{2}s_{0} & -\frac{k_{\eta}v_{+}}{\sqrt{2}}s_{0} & \frac{k_{\eta}v_{-}}{\sqrt{2}}s_{0}\\
\frac{k_{\eta}v_{+}s_{0}+a_{R,\eta}^{\dagger}}{\sqrt{2}} & -\frac{\text{\ensuremath{k_{\eta}^{\dagger}}}v_{+}}{\sqrt{2}}s_{0} & (f_{\text{BBG}}- \frac{f_{\text{TMD}}}{4})s_{0}+\frac{w_{z}}{2} & \frac{1}{2}(w_{z}+(u- \frac{f_{\text{TMD}}}{2})s_{0})\\
\frac{k_{\eta}v_{-}s_{0}+a_{R,\eta}^{\dagger}}{\sqrt{2}} & \frac{k_{\eta}^{\dagger}v_{-}}{\sqrt{2}}s_{0} & \frac{1}{2}(w_{z}+(u- \frac{f_{\text{TMD}}}{2})s_{0}) & (f_{\text{BBG}}- \frac{f_{\text{TMD}}}{4})s_{0}+\frac{w_{z}}{2}
\end{array}\right),
\end{equation}
where we denote $a_{R,\eta}=\frac{\lambda_{R}}{2}\left(\eta s_{y}+is_{x}\right)$,
$v_{\pm}=v_{0}\pm v_{4}$ and $w_{z}=\frac{1}{2}\lambda_{I}s_{z}\eta$.Let us notice that we can label eigenstates by a quantum number $m=-1,0,1$
in a next way:
\begin{equation}
H_{\gamma_{1}}\left|m\right\rangle =m\gamma_{1}\left|m\right\rangle.
\end{equation}
Our focus will be on the low energy subspace denoting $m=0$.
One can find that we can express our perturbation through ladder operators
$T_{i}\left|m\right\rangle \propto\left|m+i\right\rangle $ and $T_{i}=T_{-i}^{\dagger}$:
\begin{equation}
H^{'}=T_{-2}+T_{-1}+T_{0}+T_{1}+T_{2}+V,
\end{equation}
where:

\begin{equation}
T_{0}=\left(\begin{array}{cccc}
\frac{1}{2}(2w_{z}+(u + f_{\text{TMD}})s_{0}) & 0 & 0 & 0\\
0 & -\frac{1}{2}us_{0} & 0 & 0\\
0 & 0 & (f_{\text{BBG}}- \frac{f_{\text{TMD}}}{4})s_{0}+\frac{w_{z}}{2} & 0\\
0 & 0 & 0 & (f_{\text{BBG}}- \frac{f_{\text{TMD}}}{4})s_{0}+\frac{w_{z}}{2}
\end{array}\right),
\end{equation}
\begin{equation}
T_{1}=\frac{1}{\sqrt{2}}\left(\begin{array}{cccc}
0 & 0 & k_{\eta}^{\dagger}v_{+}s_{0}+a_{R,\eta} & 0\\
0 & 0 & -k_{\eta}v_{+}s_{0} & 0\\
0 & 0 & 0 & 0\\
k_{\eta}v_{-}s_{0}+a_{R,\eta}^{\dagger} & k_{\eta}^{\dagger}v_{-}s_{0} & 0 & 0
\end{array}\right),
\end{equation}
\begin{equation}
T_{2}=\frac{1}{2}\left(\begin{array}{cccc}
0 & 0 & 0 & 0\\
0 & 0 & 0 & 0\\
0 & 0 & 0 & 0\\
0 & 0 & (w_{z}+(u- \frac{f_{\text{TMD}}}{2})s_{0}) & 0
\end{array}\right),
\end{equation}
\begin{equation}
V=\left(\begin{array}{cccc}
0 & -k_{\eta}v_{3}s_{0} & 0 & 0\\
-k_{\eta}^{\dagger}v_{3}s_{0} & 0 & 0 & 0\\
0 & 0 & 0 & 0\\
0 & 0 & 0 & 0
\end{array}\right)=v_{3}s_{0}\left(\begin{array}{cccc}
0 & -k_{\eta} & 0 & 0\\
-k_{\eta}^{\dagger} & 0 & 0 & 0\\
0 & 0 & 0 & 0\\
0 & 0 & 0 & 0
\end{array}\right).
\end{equation}

For $H^{'}$we have the next stationary Schrödinger equation:
\begin{equation}
\left(E-H_{\gamma_{1}}-T_{-2}-T_{-1}-T_{0}-T_{1}-T_{2}-V\right)\left|\psi\right\rangle =0.
\end{equation}
From it, according to Brilluen-Wigner perturbation theory, we will receive all possible "excitation" of the lowest energy level wave function and the effective Hamiltonian that describes it.

Let's define projection to high energy states $Q=\begin{pmatrix}0 & 0 & 0 & 0\\
0 & 0 & 0 & 0\\
0 & 0 & 1 & 0\\
0 & 0 & 0 & 1
\end{pmatrix}$ and derive the iterative equation for wave-function by direct action of projection operator to Schrödinger equation:
\begin{equation}
Q\left(E-H_{\gamma_{1}}-T_{-2}-T_{-1}-T_{0}-T_{1}-T_{2}-V\right)\left|\psi\right\rangle =0,
\end{equation}
We know that $E,T_{0},V\text{ and }H_{\gamma_{1}}$ don't change our
quantum number, hence $\left[Q,E-T_{0}-V\right]=0$. Substitution
of this equality leads to the equation:
\begin{equation}
\left(E-T_{0}-V-H_{\gamma_{1}}\right)Q\left|\psi\right\rangle =Q\left(T_{-2}+T_{-1}+T_{1}+T_{2}\right)\left|\psi\right\rangle ,
\end{equation}
\begin{equation}
\label{recur1}
Q\left|\psi\right\rangle =\frac{1}{\left(E-T_{0}-V-H_{\gamma_{1}}\right)}Q\left(T_{-2}+T_{-1}+T_{1}+T_{2}\right)\left|\psi\right\rangle ,
\end{equation}
where last equality is well defined in terms of inverse operator due
to the fact that we act in a high-energy state, according to projection
operator, so the absolute value of the lowest eigenvalue $\left|\lambda\right|$
of $E-T_{0}-V-H_{\gamma_{1}}$ stay in vicinity of $\gamma_{1}$,
namely $\left|\lambda\right|\sim\gamma_{1}$. We want to find low-energy effective Hamiltonian, which corresponds to low-energy wave functions:
\begin{equation}
\left|\psi_{0}\right\rangle =\left(1-Q\right)\left|\psi\right\rangle.
\end{equation}
Inserting it into Eq.~(\ref{recur1}) we receive:
\begin{equation} \begin{split}
\left|\psi\right\rangle  & =\left|\psi_{0}\right\rangle +\frac{1}{\left(E-T_{0}-V-H_{\gamma_{1}}\right)}Q\left(T_{-2}+T_{-1}+T_{1}+T_{2}\right)\left|\psi\right\rangle, \\
 & =\left|\psi_{0}\right\rangle +\hat{A}\left|\psi\right\rangle.
\end{split} \end{equation}

So let's divide our perturbed wave function labelled by order of $\frac{1}{\gamma_{1}}$, namely $\left|\psi^{(i)}\right\rangle =...+O(\frac{1}{\gamma^{i+1}})$:
\begin{equation}
\left|\psi^{(0)}\right\rangle =\left|\psi_{0}^{(0)}\right\rangle, 
\end{equation}
\begin{equation}
\left|\psi^{(1)}\right\rangle =\left|\psi_{0}^{(1)}\right\rangle -\left(\frac{1}{H_{\gamma_{1}}}Q\right)\left(T_{1}+T_{-1}\right)\left|\psi_{0}^{(1)}\right\rangle. 
\end{equation}

The Wave function is a 4-dimensional vector on spin and zero energy $(m=0)$ spaces and we will treat it as a 2-dimensional that is block-diagonal on spin space. Consequently Effective Hamiltonian in this case will be:
\begin{equation}
\mathcal{H}^{i}_{\text{eff}}= \begin{pmatrix}
           \left\langle \psi^{(i)}_{1}\right|H\left|\psi^{(i)}_{1}\right\rangle & \left\langle \psi^{(i)}_{1}\right|H\left|\psi^{(i)}_{2}\right\rangle\\
           \left\langle \psi^{(i)}_{2}\right|H\left|\psi^{(i)}_{1}\right\rangle & \left\langle \psi^{(i)}_{2}\right|H\left|\psi^{(i)}_{2}\right\rangle 
     \end{pmatrix}.
\end{equation}

In the zeros order the effective Hamiltonian has a next form:
\begin{equation}
    \mathcal{H}^{0}_{\text{eff}}=\left(
\begin{array}{cc}
 \frac{1}{2} (2w_{z}+ (f_{\text{TMD}}+u) s_{0}) & -k_{\eta} v_{3} s_{0}\\
 -k_{\eta}^{\dagger} v_{3} s_{0} & -\frac{1}{2} u s_{0} \\
\end{array}
\right)
\end{equation}
and in first:
\begin{equation}\label{HMeff}
    \mathcal{H}^{1}_{\text{eff}}=\mathcal{H}^{0}_{\text{eff}}+\frac{1}{2\gamma_{1}}\left(
\begin{array}{cc}
 (v_{+}-v_{-}) ( |k_{\eta}|^{2} (v_{+}+v_{-})  s_{0}+ k_{\eta} a_{R,\eta}+k_{\eta}^{\dagger} a_{R,\eta}^{\dagger}) & -(v_{+}+v_{-}) \left( \frac{\left(v_{+}^2+v_{-}^2\right)}{\left(v_{+}+v_{-}\right)}(k_{\eta}^{\dagger})^{2}  s_{0}+k_{\eta}^{\dagger}a_{R,\eta} \right) \\
 -(v_{+}+v_{-}) \left( \frac{\left(v_{+}^2+v_{-}^2\right)}{\left(v_{+}+v_{-}\right)} (k_{\eta})^{2}  s_{0}+k_{\eta}a_{R,\eta}^{\dagger} \right) & (v_{+}-v_{-})|k_{\eta}|^{2}  \left(v_{+}+v_{-}\right)s_{0} \\
\end{array}
\right).
\end{equation}

One interesting phenomenon can be extracted from this Hamiltonian -- the decay of renormalized Rashba Spin-Orbit coupling in low energy subspace according to the properties of BBG. Also, that claim was independently received in the supplementary materials of the work \cite{Masseroni2024}.

\subsection*{S.2 Explicit construction of BdG Hamiltonian}
We construct the BdG Hamiltonian using the material Hamiltonian in the standard way, accounting for a subtle aspect arising due to the presence of two valleys, $\eta = \pm 1$, in a material with a hexagonal lattice. Specifically, we derive Eq.~(\ref{HMeff}) the effective low-energy Hamiltonian $\mathcal{H}_{\text{eff},\eta}$, which have zero-momentum points at the $K$ and $K^\prime$ points, respectively. To formulate the BdG model, we must return to the real zero points after diagonalizing this Hamiltonian. This can be achieved using the translation operator 
$U_{\eta} = \exp(i \eta K x)$, leading to the relation:
\begin{equation}
   \sum_{\eta}\int d\vec{k} \vec{\psi}_{\eta}^\dagger(\vec{k}) \mathcal{H}_{\text{v},\eta} \vec{\psi}_{\eta}(-\vec{k}) = 
   \sum_{\eta}\int d\vec{k}  \vec{\psi}_{\eta}^\dagger(\eta K \hat{x} + \vec{k}) \left(U_{\eta} \mathcal{H}_{\text{v},\eta} U_{\eta}^\dagger\right)  \vec{\psi}_{\eta}(\eta K \hat{x} - \vec{k}),
\end{equation}
where $\vec{k}$ is the 2D momentum vector, $\psi_{\eta}(\vec{k})$ is the annihilation operator in the $\eta$-valley, $\mathcal{H}_{\text{v},\eta}$ is the projection of effective Hamiltonian on to valence band, and the operator $U_{\eta}$ acts as:
\begin{equation}
U_{\eta} \vec{\psi}_{\eta}(\vec{k}) = \vec{\psi}_{\eta}(\eta K \hat{x} + \vec{k}).
\end{equation}
Where we emphasise that operators are 2-dimensional cause we have two spins at each valley.

To construct the BdG Hamiltonian, we define the Nambu-Gor'kov spinor basis as:
\begin{equation}
\Psi(\vec{k}) = \begin{pmatrix}
\vec{\psi}_{\eta=1}(K \hat{x} - \vec{k}) \\
\vec{\psi}_{\eta=-1}(-K \hat{x} - \vec{k}) \\
\vec{\psi}_{\eta=1}^\dagger(K \hat{x} + \vec{k}) \\
-\vec{\psi}_{\eta=-1}^\dagger(-K \hat{x} + \vec{k})
\end{pmatrix},
\end{equation}
and write the resulting BdG Hamiltonian as:
\begin{equation}
\begin{split}
    \mathcal{H}_{\text{BdG}} &= \int d\vec{k} \Psi^\dagger(\vec{k}) \left(
    \begin{array}{cccc}
    U_{+} \mathcal{H}_{\text{v},+}(\vec{k}) U_{+}^\dagger & 0 & U_{-}\hat{\Delta} U_{+} & 0 \\
    0 & U_{-} \mathcal{H}_{\text{v},-}(\vec{k}) U_{-}^\dagger & 0 & U_{+} \hat{\Delta} U_{-} \\
    U_{-}^\dagger \hat{\Delta}^* U_{+}^\dagger & 0 & -U_{+}^\dagger \mathcal{H}_{\text{v},+}^\dagger(-\vec{k}) U_{+} & 0 \\
    0 & U_{+}^\dagger \hat{\Delta}^* U_{-}^\dagger & 0 & -U_{-}^\dagger \mathcal{H}_{\text{v},-}^\dagger(-\vec{k}) U_{-}
    \end{array}
    \right) \Psi(\vec{k}).
\end{split}
\end{equation}
where $\hat{\Delta}=\Delta s_{0}$ in spin space

Simplifying further, we note the property $U_{\eta}^\dagger = U_{-\eta}$, which allows us to rewrite the Hamiltonian as:
\begin{equation}
\begin{split}
    \mathcal{H}_{\text{BdG}} &= \int d\vec{k} \Psi^\dagger \left(
    \begin{array}{cccc}
    U_{+} & 0 & 0 & 0 \\
    0 & U_{-} & 0 & 0 \\
    0 & 0 & U_{+}^\dagger & 0 \\
    0 & 0 & 0 & U_{-}^\dagger
    \end{array}
    \right)
    \left(
    \begin{array}{cccc}
    \mathcal{H}_{\text{v},+}(\vec{k}) & 0 & \hat{\Delta} & 0 \\
    0 & \mathcal{H}_{\text{v},-}(\vec{k}) & 0 & \hat{\Delta} \\
    \hat{\Delta}^* & 0 & -\mathcal{H}_{\text{v},-}^\dagger(-\vec{k}) & 0 \\
    0 & \hat{\Delta}^* & 0 & -\mathcal{H}_{\text{v},+}^\dagger(-\vec{k})
    \end{array}
    \right)
    \left(
    \begin{array}{cccc}
    U_{+}^\dagger & 0 & 0 & 0 \\
    0 & U_{-}^\dagger & 0 & 0 \\
    0 & 0 & U_{+} & 0 \\
    0 & 0 & 0 & U_{-}
    \end{array}
    \right) \Psi.
\end{split}
\end{equation}

Thus, we demonstrate that the BdG Hamiltonian in the $k = 0$ basis can be translated directly to one centred around the $K$ and $K^\prime$ points. 

\subsection*{S.3 Spectrum of spin-orbit coupled $2$JJ}


The sub-gap excitation spectrum can be directly derived in $2$JJ systems made from spin-orbit coupled van der Waals materials under the zero normal scattering limit. This simplification is feasible because, unlike 2D electron gases in semiconductors, these materials exhibit dominant Ising spin-orbit coupling for small momenta, significantly simplifying the calculations.

Let's start with the expansion of our Hamiltonian near the Fermi energy:
\begin{equation}
    H= -i v_{ {\rm f},b}(k_{x}) \partial_{y}
\end{equation}
where $v_{ {\rm f}.b}$ is momentum dependent Fermi velocity on branch $b=s,l$. Thus, we can regularly write BdG Hamiltonian:
\begin{equation}
    H= \left(\begin{array}{cc}
       -i v_{ \rm f,b}(k_{x}) \partial_{y}    & \Delta(y) \\
       \Delta^{*}(y)  & i v_{ {\rm f},b}(k_{x}) \partial_{y} 
    \end{array}\right)
\end{equation}
where 
\begin{equation}
\Delta(y)=\Delta e^{-i\phi}\Theta(W_{\text{s}}/2+W_{\text{n}}-y)+\Delta e^{i\theta}\left(\Theta(y+W_{\text{s}}/2)-\Theta(y-W_{\text{s}}/2)\right)+\Delta e^{i\phi}\Theta(y-W_{\text{s}}/2-W_{\text{n}})
\end{equation}
is the superconducting order parameter.

Matching of the boundary conditions through the junction gives us the following equation for energy:
\begin{equation}
\begin{split}
  c_{s}\left(
\begin{array}{c}
 1 \\
 \exp \left(-i \left(\phi +\nu_{b}\right)\right) \\
\end{array}
\right) &= \left(
\begin{array}{cc}
 \exp (i k_{e,b} W_{\text{n}}) & 0 \\
 0 & \exp (i k_{h,b} W_{\text{n}}) \\
\end{array}
\right)\\ &\times\left(
\begin{array}{cc}
 \exp \left(\frac{i \kappa_{-,b} W_{\text{s}}}{2}\right) & \exp \left(\frac{i \kappa_{+,b} W_{\text{s}}}{2}\right) \\
 \exp \left(i \left(\nu_{b}-\theta \right)\right) \exp \left(\frac{i \kappa_{-,b} W_{\text{s}}}{2}\right) & \exp \left(-i \left(\theta +\nu_{b}\right)\right) \exp \left(\frac{i \kappa_{+,b} W_{\text{s}}}{2}\right) \\
\end{array}
\right)\\ &\times \left(
\begin{array}{cc}
 \exp \left(-\frac{i \kappa_{-,b} W_{\text{s}}}{2} \right) & \exp \left(-\frac{i \kappa_{+,b} W_{\text{s}}}{2} \right) \\
 \exp \left(i \left(\nu_{b}-\theta \right)\right) \exp \left(-\frac{i \kappa_{-,b} W_{\text{s}}}{2} \right) & \exp \left(-i \left(\theta +\nu_{b}\right)\right) \exp \left(-\frac{i \kappa_{+,b}  W_{\text{s}}}{2} \right) \\
\end{array}
\right)^{-1}\\
&\times\left(
\begin{array}{cc}
 \exp (i k_{e,b} W_{\text{n}}) & 0 \\
 0 & \exp (i k_{h,b} W_{\text{n}}) \\
\end{array}
\right)\left(
\begin{array}{c}
 1 \\
 \exp \left(i \left(\phi +\nu_{b}\right)\right) \\
\end{array}
\right),
\end{split}
\end{equation}
where $\nu_{b}=\arccos{E/\Delta}$, $k_{e/h,b}=\pm \frac{E}{v_{ {\rm f},b}(k_{x})}$ and $\kappa_{\pm }=\pm i\frac{\sqrt{\Delta^{2}-E^{2}}}{v_{{\rm f},b}(k_{x})}$.

After some simple math transformations, one can obtain the final equation:
\begin{equation}\label{spec}
\sin\left(\nu_{b}+\phi-\left(k_{e,b}-k_{h,b}\right)2 W_{\text{n}}\right)\sin(\nu_{b})\left(-i\cot\left(\frac{1}{2}W_{\text{s}}(\kappa_{-,b}-\kappa_{+,b})\right)\right)-\cos\left(\nu_{b}+\phi-\left(k_{e,b}-k_{h,b}\right)2W_{\text{n}}\right)\cos(\nu_{b})=-\cos\left(\theta\right).
\end{equation}
\subsubsection{Solution for $\beta\neq0$}
The corresponding solution for the arbitrary angle of the pads can be obtained by direct rotation of our coordinate system:
\begin{equation}
    \begin{cases}
k_{\tilde{x}}\to\cos\left(\beta\right)k_{x}\pm\sin\left(\beta\right)(-i\partial_{y})\\
-i\partial_{\tilde{y}}\to\mp\sin\left(\beta\right)k_{x}+\cos\left(\beta\right)(-i\partial_{y})
\end{cases}
\end{equation}
where $\pm$ refers to electrons and holes, respectively.

\end{widetext}
\end{document}